\documentclass{bmcart}

\usepackage{amsthm,amsmath}
\usepackage[utf8]{inputenc} 
\usepackage{multirow,multicol} 
\usepackage{amsmath,amssymb,amsfonts} 
\usepackage{array} 
\usepackage{bm}
\usepackage{graphicx} 
\usepackage{url}

\startlocaldefs
\endlocaldefs

\begin{document}

\begin{frontmatter}

\begin{fmbox}
\dochead{Research}


\title{Weighing Features of Lung and Heart Regions for Thoracic Disease Classification}


\author[
   addressref={aff1,aff2,aff3},                   
   email={11949039@mail.sustech.edu.cn}   
]{\inits{} \fnm{Jiansheng} \snm{Fang}}
\author[
   addressref={aff4},                   
   email={ywxu@ieee.org}   
]{\inits{} \fnm{Yanwu} \snm{Xu}}
\author[
   addressref={aff4},                   
   email={yitian.zhao@nimte.ac.cn}   
]{\inits{} \fnm{Yitian} \snm{Zhao}}
\author[
   addressref={aff5},                   
   email={ygyan@maths.hku.hk}   
]{\inits{} \fnm{Yuguang} \snm{Yan}}
\author[
   addressref={aff6},                   
]{\inits{} \fnm{Junling} \snm{Liu}}
\author[
   addressref={aff2,aff4},                   
   corref={aff2},                       
   email={liuj@sustech.edu.cn}   
]{\inits{} \fnm{Jiang} \snm{Liu}}


\address[id=aff1]{
  \orgname{School of computer science and technology, Harbin Institute of Technology}, 
  \city{Harbin},                              
  \cny{China}                                    
}
\address[id=aff2]{%
  \orgname{Department of computer science and engineering, Southern University of Science and Technology}, 
  \city{Shenzhen},                              
  \cny{China}
}
\address[id=aff3]{%
  \orgname{CVTE Research}, 
  \city{Guangzhou},                              
  \cny{China}
}
\address[id=aff4]{%
  \orgname{Cixi Institute of Biomedical Engineering, Chinese Academy of Sciences}, 
  \city{Ningbo},                              
  \cny{China}
}
\address[id=aff5]{%
  \orgname{Department of Mathematics, University of Hong Kong}, 
  \city{Hongkong},                              
  \cny{China}
}
\address[id=aff6]{%
  \orgname{Guangdong Armed Police Hospital}, 
  \city{Guangzhou},                              
  \cny{China}
}



\end{fmbox}


\begin{abstractbox}

\begin{abstract} 
\parttitle{Background} Chest X-rays are the most commonly available and affordable radiological examination for screening thoracic diseases. According to the domain knowledge of screening chest X-rays, the pathological information usually lay on the lung and heart regions. However, it is costly to acquire region-level annotation in practice, and model training mainly relies on image-level class labels in a weakly supervised manner, which is highly challenging for computer-aided chest X-ray screening. To address this issue, some methods have been proposed recently to identify local regions containing pathological information, which is vital for thoracic disease classification. Inspired by this, we propose a novel deep learning framework to explore discriminative information from lung and heart regions. \parttitle{Result} We design a feature extractor equipped with a multi-scale attention module to learn global attention maps from global images. To exploit disease-specific cues effectively, we locate lung and heart regions containing pathological information by a well-trained pixel-wise segmentation model to generate binarization masks. By introducing element-wise logical AND operator on the learned global attention maps and the binarization masks, we obtain local attention maps in which pixels are $1$ for lung and heart region and $0$ for other regions. By zeroing features of non-lung and heart regions in attention maps, we can effectively exploit their disease-specific cues in lung and heart regions. Compared to existing methods fusing global and local features, we adopt feature weighting to avoid weakening visual cues unique to lung and heart regions. Our method with pixel-wise segmentation can help overcome the deviation of locating local regions. Evaluated by the benchmark split on the publicly available chest X-ray14 dataset, the comprehensive experiments show that our method achieves superior performance compared to the state-of-the-art methods. \parttitle{Conclusion} We propose a novel deep framework for the multi-label classification of thoracic diseases in chest X-ray images. The proposed network aims to effectively exploit pathological regions containing the main cues for chest X-ray screening. Our proposed network has been used in clinic screening to assist the radiologists. Chest X-ray accounts for a significant proportion of radiological examinations. It is valuable to explore more methods for improving performance.
\end{abstract}




\end{abstractbox}
%

\end{frontmatter}



\section{Background}
Chest X-ray imaging is one of the most commonly available and affordable radiological examinations for screening and clinical diagnosis. In clinical practice, diagnosing the chest X-ray images is heavily dependent on radiologists' expertise with at least years of professional experience. And this process is time-consuming and prone to subjective assessment errors \cite{brady2012discrepancy}. Hence, it is strongly desired to develop a computer-aided diagnosis system to support clinical practitioners. Many existing works using deep learning have been proposed to automatically diagnose thoracic diseases for chest X-ray images in recent years and achieve remarkable progress, such as disease classification \cite{kumar2018boosted,guan2020multi}, abnormality detection \cite{mao2020abnormality,bozorgtabar2020salad}, chest X-ray segmentation \cite{xue2020cascaded,abdulah2020lung}, disease prediction \cite{khan2020coronet,tam2020weakly}. Among various computer-aided diagnosis tasks for chest X-ray images, our work aims to address the disease classification task. The classification task is highly challenging for computer-aided screening due to the low resolution and poor specificity of chest X-ray images.

\begin{figure}[!t]
\centerline{\includegraphics[width=0.95\linewidth]{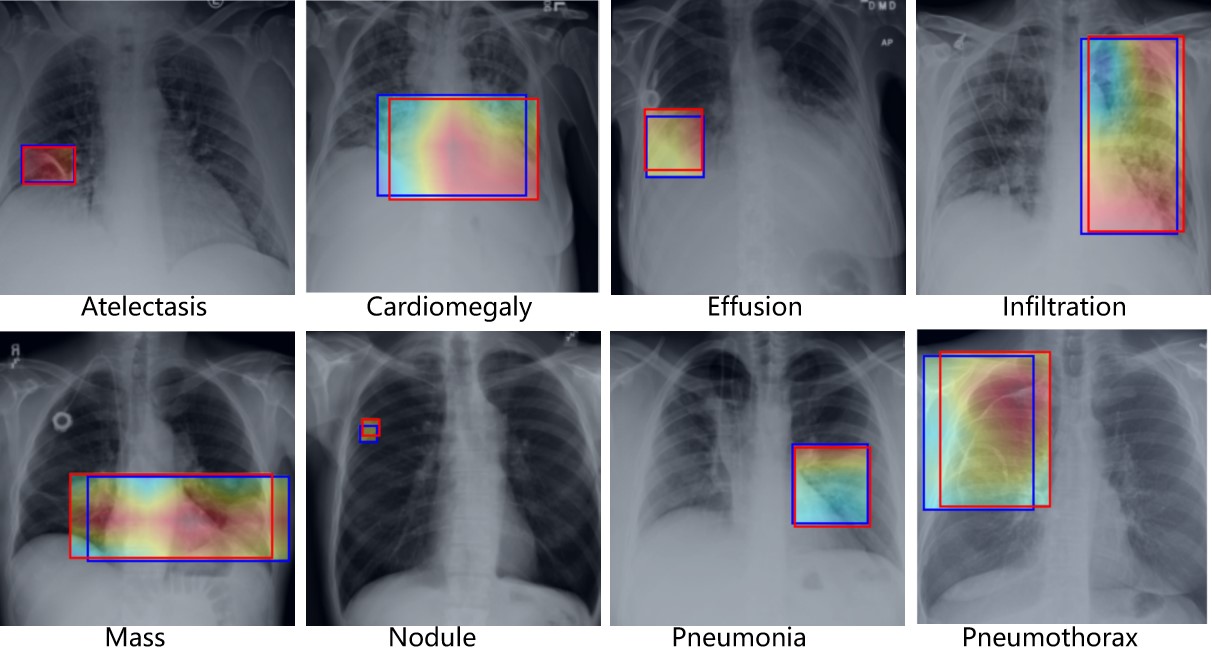}}
\caption{Examples demonstrating pathological regions of eight thoracic diseases of chest X-rays. Predicted bounding boxes by our method are shown in blue and ground truth in red.}
\label{fig1}
\end{figure}
Early works using convolutional neural networks (CNN) \cite{yao2017learning,rajpurkar2017chexnet,yan2018weakly} for thoracic disease classification of chest X-ray images typically employ the global image for model training. However, the global learning strategy may suffer from the affection of normal regions. As shown in Fig. \ref{fig1}, each image contains two parts: pathological regions (red bounding box) and normal regions. The pathological regions are the main cues for screening chest X-ray, and its cues may be drowned in the global image during model learning due to the affection of normal regions. For example, the nodule occupies a small area, and its visual cues are difficult to be reserved in the ultimate features due to a large number of convolution layers that reduce the detail characteristics. Considering this fact, it is vital to enhance the visual features of pathological regions and suppress the disturbing of normal regions during model training. However, although several large chest X-ray datasets \cite{johnson2019mimic,irvin2019chexpert,bustos2020padchest} have been published, region-level annotations are still scarce and expensive to acquire. With image-level annotations (class labels), some strategies related to pathological region locating and learning have been explored in many existing methods \cite{guan2020thorax,hermoza2020region}. 

The performance of region learning heavily relies on the accuracy of locating pathological regions with class labels. Some existing methods have been proposed to locate pathological regions for thoracic disease classification in chest X-rays, such as region proposals\cite{yao2018weakly,tang2018attention}, saliency maps\cite{yao2018weakly,tang2018attention}. However, without region-level annotations, they cannot precisely identify pathological regions by predicting bounding box, as shown in the blue rectangle of nodule image of Fig. \ref{fig1}. According to the report of existing works \cite{hermoza2020region} on the chest X-ray14 dataset \cite{wang2017chestx}, the best performance of predicting bounding box is $0.29$ average intersection over union (IoU) and $0.37$ average continuous Dice. To avoid the deviation of locating pathological regions, some works \cite{guan2020thorax,liu2019sdfn} proposed the deep fusion network by integrating the global features to compensate the lost discriminative cures of local features. However, the fusion methods must be careful tuned to avoid the local features smoothing out in the global features. The local features have learned pathological information, but its differentiating role will be weakened on the fusion process. Considering the above issues, our work designs a novel deep learning framework to explore discriminative information from local regions and enhance the differentiating role of local regions for thoracic disease classification.

By observing the area of pathological regions in Fig. \ref{fig1}, the domain knowledge that pathological regions of thoracic diseases are typically limited within the lung and heart can be asserted. Inspired by this prior knowledge, we can locate lung and heart regions by pixel-wise segmentation. Although the lung and heart regions still contain non-pathological regions that occupy large areas, these areas are smaller than the entire image and effectively cover pathological information. In fact, our method makes a trade-off between suppressing normal regions and identifying pathological regions accurately. Based on the global attention maps, the local features of the lung and heart regions are uniquely used for class-probability prediction by applying pixel-wise segmentation. Without region-level annotations, it is difficult to locate pathological regions accurately; our solution is to make the most efforts to narrow the regions containing pathological information. The main contributions of this work are summarized as follows:
\begin{itemize}
    \item[1)] To effectively learn the discriminative information from pathological regions and avoid the affection of normal regions, we propose a novel deep learning framework for thoracic diseases classification in chest X-ray. The proposed framework combines a feature extractor equipped with a multi-scale attention module and a well-trained pixel-level segmentation model for the lung and heart regions.
    \item[2)] The multi-scale attention module learns the discriminative information from chest X-ray images to generate global attention maps. We apply a feature weighting strategy for the lung and heart regions containing pathological information to exploit their disease-specific cues effectively.
    \item[3)] Evaluated by the benchmark split on the publicly available chest X-ray14 dataset, the comprehensive experiments show that our method can achieve the best performance compared to the state-of-the-art methods. The multi-scale attention module can be embedded into any off-the-shelf networks to help promote the classification performance. Our code and model have been released in \url{https://github.com/fjssharpsword/CXRDC}. 
\end{itemize}

\section{Related Works}
\textbf{Chest X-ray datasets.} 
Chest X-ray imaging is one of the most widely available modalities to assess thoracic diseases. And for a long time, the task of computer-aided screening for chest X-ray images has been extensively explored in the field of medical image analysis. Several released hospital-scale chest X-ray datasets greatly foster multi-label classification research of thoracic diseases and especially benefits the data-hungry deep learning model. For example, the MIMIC-CXR dataset \cite{johnson2019mimic} contains $377,110$ chest X-rays associated with $14$ labels, the Chexpert dataset \cite{irvin2019chexpert} provides $224,316$ chest X-rays associated with $14$ labels, the PadChest dataset \cite{bustos2020padchest} includes more than $160,000$ images labeled with $19$ differential diagnoses. Among the larger publicly available chest X-ray datasets, the Chest X-ray14 dataset \cite{wang2017chestx} attracts more research due to its earlier publish and higher quality and has been established strong baselines \cite{guan2020thorax,hermoza2020region}. Due to the comparable strong baselines, we adopt this dataset to demonstrate the advantage of our proposed method. To automatically extract the lung and heart regions from the global images, we use the JSRT dataset \cite{shiraishi2000development} to train the lung and heart segmentation model. It provides $154$ nodule and $93$ non-nodule chest X-ray images. A detailed delineation of the segmentation's nodule is publicly available to train the lung, and heart segmentation \cite{van2006segmentation}. The annotation images for segmentation tasks are binary images in which pixels are $255$ for the foreground and $0$ for the background. 

\textbf{Attention mechanisms for medical image analysis.}
Recently, attention mechanisms applied in CNN can significantly enhance the performance of various tasks in the field of medical image analysis \cite{oktay2018attention,nie2018asdnet,li2019attention}. For instance, A novel Attention Gate (AG) \cite{schlemper2019attention} can be easily integrated into standard CNN models to leverage salient regions in medical images for various medical image analysis tasks, including fetal ultrasound classification and 3D computed tomography (CT) abdominal segmentation. Attention mechanisms can help detect subtle differences between different diseases by guiding the model activations to focus on salient regions. This feature is particularly suitable for analyzing chest X-ray images due to the low resolution and poor specificity of chest X-ray images \cite{wang2019thorax,ma2019multi}. For example, a contrast-induced attention network \cite{liu2019align} is proposed to exploits the highly structured property of chest X-ray images and localizes diseases via contrastive learning on the aligned positive and negative samples. For the multi-label classification problem of thoracic diseases, an attention-guided mask inference process is designed to locate salient regions and learn the discriminative feature for classification \cite{guan2020thorax}. Inspired by this work, we improve the spatial-attention module in CBAM \cite{woo2018cbam} to design a multi-scale attention module, which helps explore discriminative cues to advance the classification performance by detecting subtle differences.

\textbf{Local Learning for chest X-ray classification.} 
Due to the relative scarcity of region-level annotations, local localization and learning are gaining increasing attention in the field of chest X-ray image analysis \cite{viniavskyi2020weakly,wolleb2020descargan}. A thoracic disease is highly characterized by a pathological region, which contains critical cues for classification. With only image-level class labels, previous works \cite{rajpurkar2017chexnet,kumar2018boosted,yao2017learning} for thoracic disease classification typically learn the discriminative information from the global image by supervised training. However, it is prone to be affected by normal regions. To address the problems caused by merely relying on the global image, recent approaches have shifted to learn the discriminative information from local regions containing pathological information. For example, a deep learning framework (SENet) \cite{yan2018weakly} equipped with the squeeze-and-excitation block \cite{hu2018squeeze} reinforces the sensitivity to subtle differences between normal and pathological regions by explicitly modeling the channel interdependence. More methods for local location rely on saliency maps or saliency maps \cite{yao2018weakly,tang2018attention,hermoza2020region}. For instance, in SalNet \cite{hermoza2020region}, the Gumbel-softmax function \cite{jang2016categorical} is used to combine the region proposal and saliency map detector to sample discrete regions from a set of proposed regions differentially. However, without region-level annotations, they cannot precisely identify pathological regions by selecting local regions. 

To avoid discriminative information loss in location deviation of pathological regions, some methods fuse the global image training and the local region learning. The deep fusion network unifying global and local features is gradually popular in computer vision tasks \cite{ding2017local,cao2020unifying}. For thoracic disease classification in chest X-ray images, the representative work of fusion methods are the segmentation-based deep fusion network (SDFN) \cite{liu2019sdfn} and the three-branch attention-guided network (AGCNN) \cite{guan2020thorax}. In SDFN, a global classifier is used as feature extractors to obtain the discriminative features from the entire chest X-ray image, and the cropped lung regions generated by the segmentation model are learned by a local classifier. The obtained features from the global and local classifiers are fused by the feature fusion module for disease classification. Our method and SDFN all use the JSRT dataset \cite{van2006segmentation} to train a pixel-wise segmentation model. However, the fusion methods must be careful tuned to prevent the local features containing pathological information from drowning in the global features. Hence, we apply feature weighting but not fusion to enhance visual cues unique to the lung and heart regions based on the learned global attention maps and the segmented masks. 

Based on the above discussion of related works, our proposed method has two novel folds: (1) a feature extractor equipped with the multi-scale attention module is used to learn the global discriminative information; (2) feature weighting strategy is applied to enhance features of the lung and heart region containing pathological information. Extensive experiments on the chest X-ray14 dataset demonstrate the effectiveness of our method.

\section{Methods}
Based on image-level class labels, our method is proposed to address the multi-label classification of thoracic diseases by learning the discriminative information from chest X-ray images effectively. This section will elaborate on our method, including the problem statement, feature extractor, feature weighting.

\subsection{Problem Statement}
\begin{figure*}[!t]
\centerline{\includegraphics[width=0.95\linewidth]{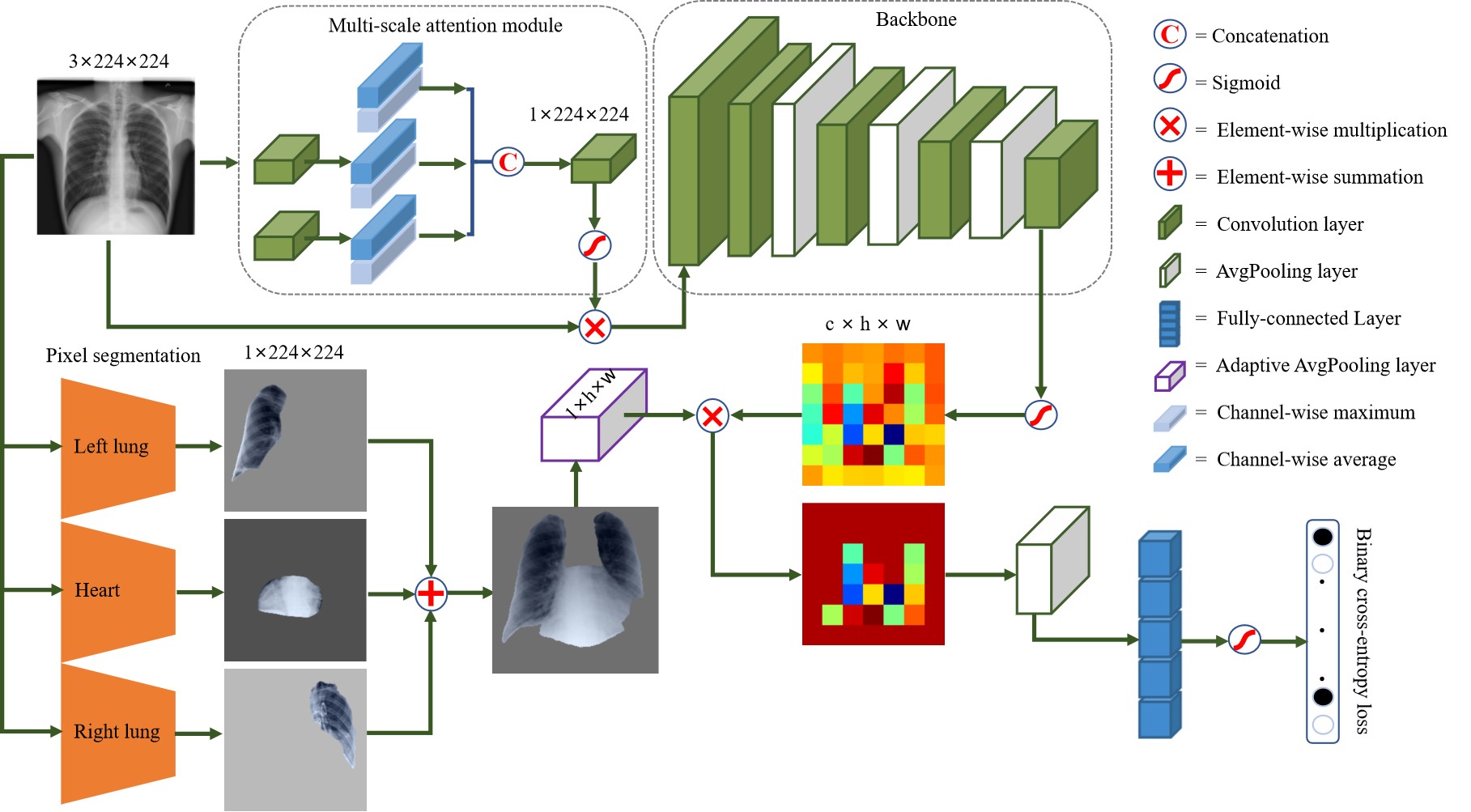}}
\caption{The framework of our proposed method. A feature extractor equipped with a multi-scale attention module aims to learn the discriminative information from a chest X-ray image to generate a global attention map. A well-trained pixel segmentation model locates the lung and heart regions to binarize a mask in which pixels are $1$ for lung and heart regions and $0$ for other regions. A local attention map focusing on the lung and heart regions is formed by introducing a logical AND operator on the mask and the global attention map. This local attention map contains features of the pathological region and suppresses the normal region.}
\label{fig2}
\end{figure*}

Thoracic disease classification is a multi-label classification problem that detects if one or multiple diseases are presented in each chest X-ray image. We define a $14$-dimensional label vector $\bm{Y}=\{y_{1}, \dots, y_{i}, \dots, y_{c}\}$ for each image, where $c=14$ and $y_{i} \in \{0, 1\}$. $y_{i}$ indicates the presence with respect to corresponding diseases in the image (\textit{i.e.} $1$ for presence and $0$ for absence) and an all-zero vector of $14$-dimensions represents the status of "No Finding" (no disease is found in the scope of any of $14$ disease categories as listed). The diseases in $\bm{Y}$ are in the order of Atelectasis, Cardiomegaly, Effusion, Infiltration, Mass, Nodule, Pneumonia, Pneumothorax, Consolidation, Edema, Emphysema, Fibrosis, Pleural Thickening, and Hernia. We address this classification problem by training our classification model presented in Fig. \ref{fig2} with the binary cross-entropy (BCE) loss function defined in Eq \ref{eq1}.
\begin{equation}
\begin{aligned} 
    BCE(\bm{Y},\bm{\hat{Y}}) = - \frac{1}{c}\sum_{i=1}^{c}[y_{i}\log(\hat{y_{i}}) + (1-y_{i})\log(1-\hat{y_{i}})]
\end{aligned}
\label{eq1}, 
\end{equation}
where $c$ is the number of diseases (classes), $\bm{Y}$ is the ground truth, and $\bm{\hat{Y}}$ denotes the predicted probability. 

Our proposed deep framework covers three parts: a feature extractor, a pixel-wise segmentation model, and a feature weighting module. The feature extractor is to embed the global discriminative information into a global attention map by applying a multi-scale attention module. The multi-scale attention module helps the feature extractor to focus on salient regions and detect subtle texture abnormality. Simultaneously, the well-trained pixel segmentation model identifies areas of the lung and heart, following binarized as a global mask in which pixels are $1$ for lung and heart region and $0$ for other regions. Then we conduct an element-wise summation operation on the global attention map and the global mask to generate a local attention map. By weighing the lung and heart region features, the local attention map only contains visual cues unique to the lung and heart region containing pathological information and discards features of non-lung and heart regions by zeroing operation. Following the local attention map, an average pooling layer and a fully-connected layer are introduced to train disease-specific probability by binary cross-entropy loss. 

\subsection{Feature Extractor}
The feature extractor consists of a multi-scale attention module and a backbone. Each chest X-ray image $\bm{X}$ is resized into $3\times 224\times 224 $ and firstly inputted into the multi-scale attention module. The multi-scale attention module computes a spatial feature hierarchy consisting of two convolutional layers with a kernel step of $2$ and three blocks of calculating maximum and average across channels. The spatial feature hierarchy is convoluted into a feature map of $1\times 224\times 224 $ dimension and merged into the global image by element-wise multiplication. Based on this operation, the global image element is spatially weighted by computing the maximum value at different scales. The multi-scale spatial attention module can detect subtle differences at different scales. Hence, it can enhance the multi-label classification performance by exploiting the visual cues effectively. After a sigmoid activation function, the feature map is merged into the original chest X-ray image by element-wise multiplication, following fed into the backbone. We use the pre-trained $121$-layer DenseNet \cite{huang2017densely} as the backbone. We take out the last convolutional feature map from backbone as a global attention map $\bm{F_{g}}$ with $c\times h\times w $ dimensions. The global attention map learns the discriminative information from the chest X-ray image. The disease-specific feature may be drowned in global features and can not play a differentiating role in classification.

\subsection{feature weighting}
We apply U-Net \cite{ronneberger2015u} to train a segmentation model for the left lung, right lung, and heart on the JSRT dataset by using dice loss. The dice loss is formulated as:
\begin{equation}
    \begin{aligned} 
        \mathrm{dice} = \frac{2|M_{gt} \bigcap M_{prob}|}{(|M_{gt}| + |M_{prob}|)}
    \end{aligned}
    \label{eq2},
\end{equation}
where $M_{gt}$ denotes the ground truth mask, and $M_{prob}$ is the predicted mask. The dice loss is minimized for optimization and the model with the smallest loss was saved. The image pre-processing of U-Net follows the same pipeline of the feature extractor to enable automatic region segmentation for the chest X-ray14 dataset. We first input the chest X-ray image $\bm{X}$ into the well-trained segmentation model to generate three pixel-wise masks for the left lung, right lung, and heart. Then we merge the three pixel-wise masks into a pixel-wise mask $\bm{M_{g}}$ in which pixels are either 1 for the lung and heart regions or 0 for other regions by pixel-wise summation. The pixel-wise mask $\bm{M_{g}}$ further is resized into a size of $ 1\times h\times w $ equal to the width and height of the global attention map $\bm{F_{g}}$ by adaptive average pooling. The global attention map $\bm{F_{g}}$ of $c\times h\times w $ is taken out from the backbone of the image classifier. Further, we generate a local attention map $\bm{F_{l}}$ of $c\times h\times w $ from the global attention map and the pixel-wise mask by element-wise multiplication. We introduce the logical AND operator on the global attention map and the pixel-wise mask. The local attention map contains the zero pixels of non-lung and heart regions and the non-zero pixels of the lung and heart regions. Hence, only the pixel values of the lung and heart region containing pathological information in the local attention map are embedded into the average pooling layer for label prediction by a channel-wise average operation, and the pixel values of other regions in the attention map are zeroed. The feature weighting for the global attention map $\bm{F_{g}}$ and the pixel-wise mask $\bm{M_{g}}$ is defined as:
\begin{equation}
    \begin{aligned} 
        \bm{F_{l}} = \bm{F_{g}} \otimes \bm{M_{g}}
    \end{aligned}
    \label{eq3}. 
\end{equation}

With the help of the multi-scale attention module, the global attention map effectively learns the salient information from the chest X-ray image, containing the discriminative information in the lung and heart. The pathological regions are typically located in the lung and heart, hence, we introduce the binary masks on the global attention map to generate the local attention map. The generated local attention map suppresses the information of other regions and remains the information of the lung and heart regions. By logical AND operation, we locate features of the lung and heart regions containing pathological information.

\section{Experimental Setups}
In order to test the performance of our proposed framework, we conduct extensive experiments on the public chest X-ray14 dataset to verify the effectiveness of our method. In this section, we will describe the experimental details. 

\textbf{Chest X-ray14 dataset} consists of $112,120$ frontal-view X-ray images of $30,805$ unique patients \cite{wang2017chestx}. Each image is labeled with one or multiple classes of $14$ common thoracic disease: Atelectasis, Cardiomegaly, Effusion, Infiltration, Mass, Nodule, Pneumonia, Pneumothorax, Consolidation, Edema, Emphysema, Fibrosis, Pleural Thickening, and Hernia. Besides, the dataset also contains $984$ labeled bounding boxes for $880$ images related to $8$ different diseases by board-certified radiologists. In our experiments, we use disease labels as ground-truth for model training. At the same time, we utilize the bounding boxes for qualitative observation of pathological region localization on chest X-rays. As Table \ref{tb1} shows, the benchmark split of this dataset \cite{wang2017chestx} contains train set of $86,524$ images for model training, test set of $25,596$ images for model evaluation, and box set of $984$ images for model visualization. We randomly select $10\%$ of each disease in the train set as the validation set for model validation. There is no patient overlap between the three splits. There are some images with multi-label, so the number of multi-label totals is greater than the finding number.
\begin{table}[htbp]
\renewcommand{\arraystretch}{1.2}
\caption{The statistics of the benchmark split on the Chest X-ray14 dataset.}
\begin{center}
    \begin{tabular}{|c|c|c|c|c|}
    \hline 
    \textbf{Diseases} & \textbf{Train} & \textbf{Test} & \textbf{Box} \\
    \hline
    \textbf{Atelectasis} & 8,280 & 3,279 & 180 \\
    \textbf{Cardiomegaly} & 1,707 & 1,069 & 146 \\
    \textbf{Effusion} & 8,659 & 4,658 & 153 \\
    \textbf{Infiltration} & 13,782 & 6,112 & 123 \\
    \textbf{Mass} & 4,034 & 1,748 & 85 \\
    \textbf{Nodule} & 4,708 & 1,623 & 79 \\
    \textbf{Pneumonia} & 876 & 555 & 120 \\
    \textbf{Pneumothorax} & 2,637 & 2,665 & 98 \\
    \textbf{Consolidation} & 2,852 & 1,815 & 0 \\
    \textbf{Edema} & 1,378 & 925 & 0 \\
    \textbf{Emphysema} & 1,423 & 1,093 & 0 \\
    \textbf{Fibrosis} & 1,251 & 435 & 0 \\
    \textbf{Pleural Thickening} & 2,242 & 1,143 & 0 \\
    \textbf{Hernia} & 141 & 86 & 0 \\
    \textbf{Multi-label Totals} & 53,970 & 27,206 & 984 \\
    \hline
    \textbf{Finding} & 36,024 & 15,735  & 984 \\
    \textbf{No Finding} & 50,500 & 9,861  & 0 \\
    \textbf{Totals} & 86,524& 25,596  & 984 \\
    \hline
    \end{tabular}
\end{center}\label{tb1}
\end{table}

\textbf{Comparative methods.} Researches on addressing the multi-label classification problem of thoracic diseases have established strong baselines on the benchmark split of the chest X-ray14 dataset. 
\begin{itemize}
    \item \textbf{DCNN \cite{wang2017chestx}.} In this work \cite{wang2017chestx}, they first released the benchmark split of the chest X-ray14 dataset and presented a deep convolutional neural network (DCNN) to tackle thoracic disease classification. We reproduce this method by using the pre-trained ResNet-50 \cite{he2016deep}, which achieved the best performance in this work.
    \item \textbf{CheXNet \cite{rajpurkar2017chexnet}.} CheXNet \cite{rajpurkar2017chexnet} is a $121$-layer DenseNet \cite{huang2017densely} trained on the chest X-ray14 datset. This work demonstrated that the performance of CheXNet is statistically significantly higher than radiologist performance. 
    \item \textbf{SENet \cite{yan2018weakly}.} To deal with the challenge that thoracic diseases usually happen in localized disease-specific areas, Yan \textit{et al.} \cite{yan2018weakly} presented a weakly-supervised deep learning framework equipped with squeeze-and-excitation blocks (SENet) to classify thoracic disease. This work was based on the CheXNet model using DenseNet as the backbone and first explored the problem of learning disease-specific areas. 
    \item \textbf{SDFN \cite{liu2019sdfn}.} Liu \textit{et al.} \cite{liu2019sdfn} provided a segmentation-based deep fusion network (SDFN) to leverage the discriminative information of local regions. SDFN adopted pixel-level segmentation to detect local regions and applied a deep fusion framework to unify the global and local features. Our method also identifies the lung and heart region by using pixel segmentation. But we argue that the deep fusion method can not effectively tackle the problem that the local features are drowned in the global features. Hence, we use a feature weighting strategy to focus on the local features.
    \item \textbf{AGCNN \cite{guan2020thorax}.} Guan \textit{et al.} proposed a three-branch attention-guided convolutional neural network (AGCNN) \cite{guan2020thorax} for the task of thoracic disease classification on chest X-ray images. This work located salient regions from the global attention map then cropped the corresponding regions from the chest X-ray image.
    \item \textbf{SalNet \cite{hermoza2020region}.} Hermoze \textit{et al.} \cite{hermoza2020region} designed a three-stage deep learning framework (SalNet) for weakly-supervised disease classification by combining region proposal and saliency detection. This work obtained the local regions from salient maps based on region proposals and achieved the best performance on the benchmark split of the chest X-ray14 dataset.
\end{itemize}

\textbf{Implementation details and evaluation protocal.} We implement CXR-IRNet with the Pytorch framework and use the pre-trained $121$-layer DenseNet as the backbone of the feature extractor. We extract the last convolutional feature map of DenseNet as the global attention map. The single output is used for class-probability prediction after a sigmoid non-linearity. For the multi-scale attention module, apart from the original image as one feature, we adopt two convolutions of kernel size $5, 9$ to generate the other two features, these three-scale features for following operations. We resize each chest X-ray image to $256 \times 256$, and then perform center cropping to obtain an image of size $224 \times 224$ for training. Each cropped image is normalized with the same mean and standard deviation. We use Adam optimizer with a learning rate of $0.001$ and weight decay of $0.0001$. Our network is trained for $50$ epochs from scratch with a batch size of $512$. For comparative methods, we directly report the published performance of SDFN and SalNet, no reproduction. The other methods are implemented by the same experimental setup for a fair comparison. For evaluation, we report the area under the receiver operating characteristic curve (AUROC) and ROC curve. Both are widely used for performance assessment of multi-label classification. The ROC curve comprises of two evaluation criteria to measure performance, including sensitivity (true positive rate) and specificity (true negative rate). For detection visualization, we evaluate in terms of the intersection over union (IoU) on the box set.

\section{Results and Discussions}
The following research questions will be answered by analyzing experimental results: 
\begin{description}
\item[\textbf{RQ1}] Can feature weighting of the lung and heart regions help improve the performance?
\item[\textbf{RQ2}] How is the effectiveness of the multi-scale attention module on learning pathological information? 
\end{description}

\subsection{Classification Performance (RQ1)}
\begin{table}[!t]
\renewcommand{\arraystretch}{1.2}
\caption{Comparison of AUROC performance on the benchmark split of Chest X-ray14 dataset.}
\begin{center}
    \begin{tabular}{|c|cccc|}
    \hline 
    \textbf{Diseases}
    & \textbf{DCNN \cite{wang2017chestx}} &\textbf{CheXNet \cite{rajpurkar2017chexnet}} 
    &\textbf{SENet \cite{yan2018weakly}} & \textbf{SDFN \cite{liu2019sdfn}}\\
    \hline
    \textbf{Atelectasis} & 0.7837 & 0.7919 & 0.7963 & 0.7810  \\
    \textbf{Cardiomegaly} & 0.8937 & 0.9038 & 0.9085 & 0.8850 \\
    \textbf{Effusion} & 0.8704 & 0.8744 & 0.8769 & 0.8320 \\
    \textbf{Infiltration} & 0.6826 & 0.6942 & 0.6974 & 0.7000 \\
    \textbf{Mass} & 0.7875 & 0.8142 & 0.8144 & 0.8150 \\
    \textbf{Nodule} & 0.7125 & 0.7286 & 0.7509 & 0.7650 \\
    \textbf{Pneumonia} & 0.7110 & 0.7477 & \textbf{0.7566} & 0.7190 \\
    \textbf{Pneumothorax} & 0.8232 & 0.8431 & 0.8467 & 0.8660 \\
    \textbf{Consolidation} & 0.7895 & 0.7933 & 0.7996 & 0.7430 \\
    \textbf{Edema} & 0.8673 & 0.8777 & \textbf{0.8864} & 0.8420 \\
    \textbf{Emphysema} & 0.8398 & 0.8726 & 0.8907 & 0.9210 \\
    \textbf{Fibrosis} & 0.7656 & 0.7986 & 0.8041 & 0.8350 \\
    \textbf{Pleural Thickening} & 0.7398 & 0.7528 & 0.7547 & 0.7910\\
    \textbf{Hernia} & 0.8406 & 0.8545 & 0.9035 & 0.9110 \\
    \textbf{Average} & 0.7934 & 0.8105 & 0.8205 & 0.8150 \\
    \hline
    \textbf{Diseases}
    & \textbf{AGCNN \cite{guan2020thorax}} & \textbf{SalNet \cite{hermoza2020region}}
    & \textbf{Ours w/o $\bm{F_{l}}$} & \textbf{Ours} \\
    \hline
    \textbf{Atelectasis}  & 0.8076 & 0.7750 & \textbf{0.8151} & 0.8073  \\
    \textbf{Cardiomegaly}  & 0.8987 & 0.8810 & \textbf{0.9100} & 0.9069 \\
    \textbf{Effusion} & 0.8682 & 0.8310 & \textbf{0.8779} & 0.8401 \\
    \textbf{Infiltration} & 0.6933 & 0.6950 & 0.6923 & \textbf{0.7901}\\
    \textbf{Mass} & \textbf{0.8314} & 0.8260 & 0.8290 & 0.8131 \\
    \textbf{Nodule} & 0.7661 & 0.7890 & 0.7553 & \textbf{0.8377} \\
    \textbf{Pneumonia} & 0.7383 & 0.7410 & 0.7401 & 0.7465 \\
    \textbf{Pneumothorax}  & 0.8431 & \textbf{0.8790} & 0.8516 & 0.8728 \\
    \textbf{Consolidation} & 0.7945 & 0.7470 & \textbf{0.8038} & 0.7831 \\
    \textbf{Edema} & 0.8791 & 0.8460 & 0.8858 & 0.8575 \\
    \textbf{Emphysema} & 0.8967 & \textbf{0.9360} & 0.8849 & 0.8605 \\
    \textbf{Fibrosis} & 0.7936 & 0.8330 & 0.8077 & \textbf{0.8494} \\
    \textbf{Pleural Thickening} & 0.7642& 0.7930 & 0.7573 & \textbf{0.8123} \\
    \textbf{Hernia} & \textbf{0.9344} & 0.9170 & 0.8995 &  0.8915 \\
    \textbf{Average} & 0.8221 & 0.8210 & 0.8222 & \textbf{0.8335} \\
    \hline
    \end{tabular}
\end{center}\label{tb2}
\end{table}
In Table \ref{tb2}, we report the classification performances of the proposed method and comparative methods in terms of AUROC scores, evaluated by the test set of the benchmark split. Our method achieves the best performance (boldface font) over $4$ diseases, including Infiltration, Nodule, Fibrosis, and Pleural Thickening. In terms of the average AUROC, our method is superior to comparative methods. The overall results show that our method establishes a new state-of-the-art on the benchmark split of the chest X-ray14 dataset. Methods (SENet, SDFN, AGCNN, SalNet) unifying the global and local features obtain better performance than methods (DCNN, CheXNet) only employing the global image. To overcome location deviation of methods (AGCNN, SalNet) relying on saliency maps and region proposal, our method identifies the lung and heart regions containing pathological information by using pixel-wise segmentation same as SDFN. However, we argue that unifying the global and local features can not prevent local discriminative information from smoothing out in the global features. Hence, we consider the feature weighting strategy but not fusion like SDFN and AGCNN. SENet locates suspicious lesion regions by using a multi-map transfer layer to encode activations associated with each disease class. Such feature weighting strategy makes it more capable of discriminating the appearance of multiple thoracic diseases on the same chest X-ray, then helps it yields good performance. Different from SENet, Our method conduct feature weighting on the global attention map by using segmentation masks. Benefit from the segmentation locating the lung and heart regions containing pathological information precisely and the feature weighting strategy zeroing features of non-lung and heart regions in attention maps, our method establish a new baseline on the chest X-ray14 dataset.

As Table \ref{tb2} show, without feature weighting (\textbf{Ours w/o $\bm{F_{l}}$}), our deep framework equipped with the multi-scale attention module can achieve the competitive performance, including the highest performance of $4$ diseases and the second-highest average performance. Without feature weighting, our deep framework is equal to ChXNet employing the global image. The improved performance demonstrates that the multi-scale attention module can effectively learn the discriminative information from the global image. Only the discriminative information is learned into the global attention maps, the feature weighting on the global attention maps can locate the lung and heart regions containing pathological information. The multi-scale attention module exploits the salient information from chest X-ray at three scales, then detects visual cues unique to pathological regions. The performances of \textbf{Ours w/o $\bm{F_{l}}$} confirm the contribution of the multi-scale attention module. With the help of the multi-scale attention module, our method can further improve the performance by applying the feature weighting strategy to enhance the visual cues unique to the lung and heart regions. We argue that local features containing pathological information maybe drown in the global feature by applying the fusion framework like AGCNN and SDFN. Hence, based on locating the lung and heart regions containing pathological information by segmentation, we directly zero features of non-lung and heart regions.

To further demonstrate the advantage of the feature weighting strategy, we can observe the performance of some diseases. The Infiltration AUROC of our method is significantly improved compared to other methods. The improvement ratio reaches $12.87\%$ compared to the second-highest performance yielded by SDFN. As Table \ref{tb1} show, the number of Infiltration image is the most among the diseases. However, other methods can not achieve better performance due to the poor specificity of Infiltration. At the same time, we can observe that the pathological region of Infiltration occupies a relatively large area in the left lung, as shown in Fig. \ref{fig1}. This demonstrates that the effectiveness of the feature weighting strategy. The pathological region of Infiltration covers a large area of the left lung, and features of the left lung are enhanced after feature weighting. Hence, the class-probability prediction mainly relies on the learned pathological information of Infiltration. In other words, the pathological information of Infiltration is not weakened or even lost in the pipeline of our deep framework, while the non-pathological regions are suppressed. Benefiting from weighing features of the lung and heart regions, the performance of Nodule is up to $0.8377$ obtained by our method. The pathological region of Nodule is usually small and easily drowned in the global image. The characteristics and performance of Nodule also demonstrate the effectiveness of the feature weighting strategy zeroing features of non-pathological regions. Based on the above discussion, we can infer that the feature weighting strategy can help improve classification performance and is superior to the fusion method.

\begin{figure}[!t]
\centerline{\includegraphics[width=\columnwidth]{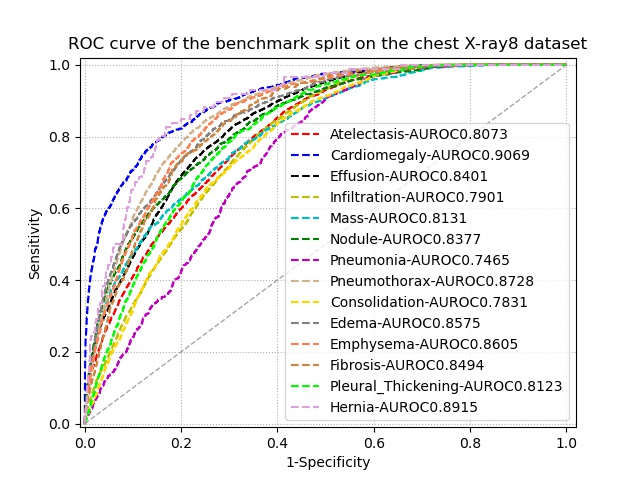}}
\caption{ROC Curves of our proposed method on the benchmark split of Chest X-ray14 dataset.}
\label{fig3}
\end{figure}
Fig. \ref{fig3} shows the ROC curves of our method on the $14$ diseases of the benchmark split. According to the ROC curve trained on the chest X-ray14 dataset, we set the class threshold for each disease to classify a new chest X-ray image. Due to the reliable performance, our model has been successfully applied in routine clinical screening to assist radiologists \footnote{http://www.yibicom.com/}. We automatically output screening results of our method before the radiologists read the chest X-ray images in the picture archiving and communication systems (PACS). On the user interface of PACS, the radiologists can get the pre-screened result to make further diagnosis. For automatic screening chest X-rays, the underlying idea is to effectively suppress non-pathological regions and learn visual cues of pathological regions. In this work, we devote ourselves to locating the lung and heart regions containing pathological information by designing the multi-scale attention module and feature weighting strategy. Our proposed framework can avoid the deviation in locating pathological regions by using pixel-wise segmentation and the local features drown in the global features by using feature weighting. In the future, we try to improve our model by applying region-wise detection to learn visual cues unique to pathological regions. 

\subsection{Learning Capability (RQ2)}
\begin{figure}[!t]
\centerline{\includegraphics[width=0.95\columnwidth]{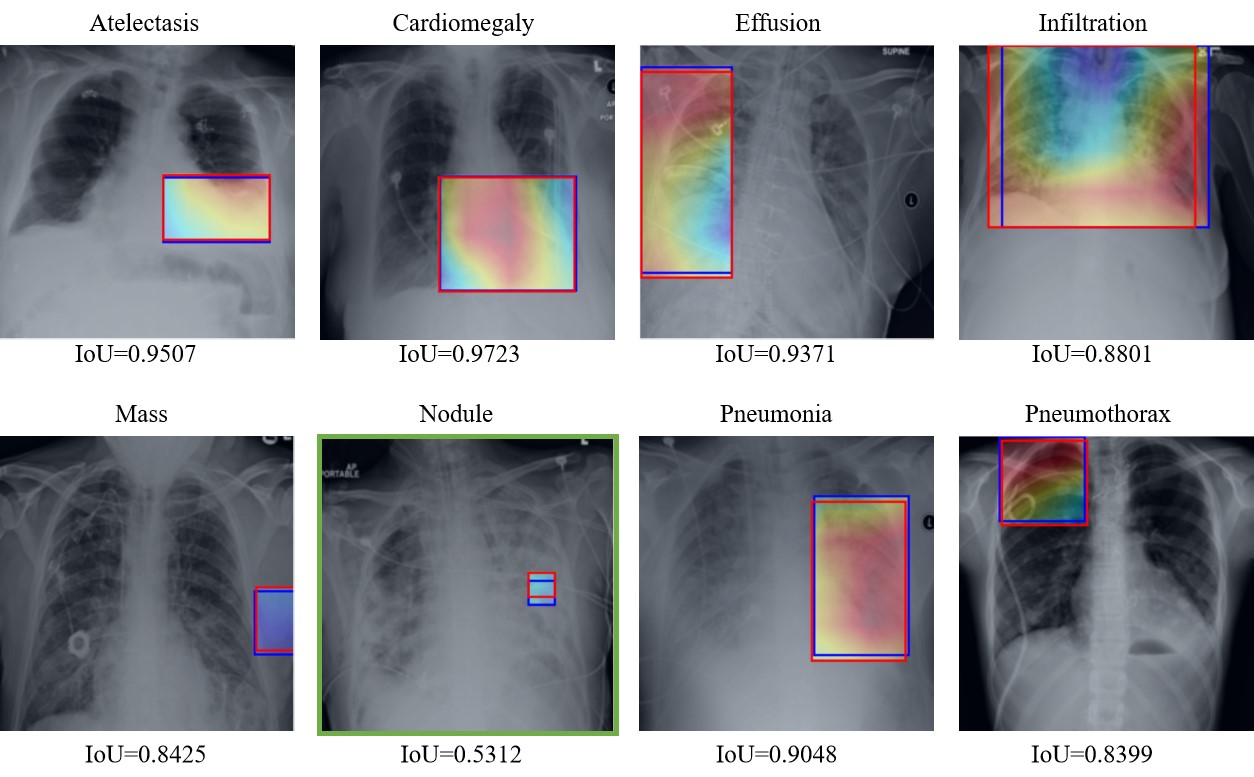}}
\caption{Pathological region visualization of the box set in the Chest X-ray14 dataset. The red bounding box is the pathological region by hand-labeled, and the blue bounding box is predicted by our model.}
\label{fig4}
\end{figure}
\begin{table*}[!t]
\renewcommand{\arraystretch}{1.2}
\caption{Comparison of IoU performance on the box set of the chest X-ray14 dataset.}
\begin{center}
    \begin{tabular}{|c|c|c|c|c|}
    \hline 
    \textbf{Methods} & \textbf{Atelectasis} & \textbf{Cardiomegaly} & \textbf{Effusion} & \textbf{Infiltration}  \\
    \hline
    \textbf{FE} & 0.1713 & 0.5515 & 0.3242 & 0.3819 \\
    \textbf{FE w/o MA} & 0.1212 & 0.4430 & 0.3258 & 0.2783 \\
    \hline
    \textbf{Methods} & \textbf{Mass} & \textbf{Nodule} & \textbf{Pneumonia} & \textbf{Pneumothorax} \\
    \hline
    \textbf{FE} & 0.1768 & 0.0460 & 0.3097 & 0.3345 \\
    \textbf{FE w/o MA} & 0.1429 & 0.0418 & 0.2328 & 0.2640 \\
    \hline
    \end{tabular}
\end{center}\label{tb3}
\end{table*}
The capability of learning pathological information determines the final classification performance. Even if the lung and heart regions can be located accurately by pixel-wise segmentation, but if the feature extractor can not learn the pathological information in the lung and heart regions, the feature weighting strategy can not help improve the performance. So we need to analyze the effectiveness of the multi-scale attention module in learning pathological information. The best average AUROC in Table \ref {tb2} can demonstrate that our proposed method has reliable learning capability. Apart from this proof, we further adopt the box set with ground truth (bounding box) to evaluate the learning capability of pathological information of the feature extractor equipped with the multi-scale attention module. We apply class activation map (CAM) \cite{zhou2016learning} to locate regions containing pathological information. Then we use IoU to evaluate the performance of the predicted pathological regions based on the ground truth pathological regions.

Some images with the higher IoU performance are shown in Fig. \ref{fig4}. This qualitative visualization demonstrates that the feature extractor can detect pathological regions with some probability. The detection performance can reflect the learning capability of the feature extractor equipped with the multi-scale attention module. The detection performance for Nodule (green rectangle) is lower than other diseases due to its small area, but the detected pathological region lay on the left lung. By filtering out the non-lung and heart regions, pathological information in the left lung can be used for label prediction. The detected pathological region of Cardiomegaly is almost overlapping with the heart region. The detected pathological region of Pneumonia also almost covers the left lung region. But the pathological region of Mass is severely deviating to the lung and heart region. Although the feature extractor has learned the pathological information, the pathological region will be filtered out in the process of feature weighting. Such cases affect the classification performance of our method and can be overcome by using region-level annotations. Our proposed method aims to improve the performance with image-level class labels. Further, we present an ablation study to demonstrate the contribution of the multi-scale attention module (MA) in the feature extractor (FE). In Table \ref{tb3}, the average IoU of FE can greatly outperform FE without MA by $23.67\%$ from $0.2437$ to $0.3014$. This IoU performance is competitive to SalNet that reports an average of IoU of $0.29$. Our feature extractor adopts the same backbone as DCNN, and the AUROC performance of our method without feature weighting (\textbf{Ours w/o $\bm{F_{l}}$}) is superior to DCNN in Table \ref{tb2}. Based on the above observations, we can conclude that the multi-scale attention module contributes to pathological information learning and classification performance improvement.

We typically divide a chest X-ray image into two parts: pathological region and non-pathological region. Our method aims to filter out the information of the non-pathological region. However, it is difficult to locate the pathological region without region-level annotations. Current works relying on saliency map or region proposal lead to location deviation. To overcome this issue, we apply pixel-wise segmentation to locate the lung and heart regions containing pathological information. Although the lung and heart regions can cover the pathological region in most cases, the non-pathological region in the lung and heart regions can not be filtered out by feature weighting. The feature weighting strategy only can filter out non-lung and heart regions. Despite this, our method applying the feature weighting strategy achieves better performance than methods using fusion strategy. With image-level class labels, we design two tricks to improve the performance of multi-label classification for screening chest X-rays. Based on the above experimental results and discussion, we have demonstrated the effectiveness of these two tricks. 

\section{Conclusions}
In this work, we propose a novel deep framework for the multi-label classification of thoracic diseases in chest X-ray images. The proposed network aims to effectively exploit pathological regions containing the main cues for chest X-ray screening. We present a feature extractor equipped with a multi-scale attention module to effectively learn pathological information from chest X-ray images. At the same time, we apply the pixel-level segmentation to identify the lung and heart regions containing pathological information to overcome location deviation. Then, we adopt the feature weighting strategy to filter out the non-lung and heart regions. Based on our deep framework, the class-probability layer mainly rely on the information of the lung and heart regions. Evaluated on the benchmark split of the chest X-ray14 dataset, we establish a new state-of-the-art baseline. Our proposed network has been used in clinic screening to assist the radiologists. Chest X-ray accounts for a significant proportion of radiological examinations. It is valuable to explore more methods for improving performance.


\begin{backmatter}
\section*{Ethics approval and consent to participate}
\section*{Consent for publication}

\section*{Competing interests}
The authors declare that they have no competing interests.

\section*{Funding}
This work was supported in part by the Science and Technology Innovation Committee of Shenzhen City (JCYJ20200109140820699 and 20200925174052004).

\section*{Authors' contributions}
JSF conceived the topic of this research and did the thoracic disease classification and wrote the manuscript. YWX, YTZ, YGY, and JLL participated in its design and revised it critically for the important intellectual content. JL conceived of the survey and participated in designing it. All authors read and approved the ﬁnal manuscript.

\section*{Acknowledgements}
The authors would like to thank many members of the Intelligent Medical Imaging (iMED) group for the inspiring knowledge sharing, technical discussions, clinical background infusion. 

\section*{Availability of data and materials}
\section*{Abbreviations}
\section*{Authors' information}
\section*{Data Availability Statement}
Not applicable.
\section*{Ethical Declaration Statement}
NA.

\bibliographystyle{bmc-mathphys} 
\bibliography{CXRNet}      
\end{backmatter}
\end{document}